\documentclass[10pt,conference]{IEEEtran}
\makeatletter
\renewcommand{\@IEEEsectpunct}{~}
\makeatother

\ifCLASSINFOpdf
  \usepackage[pdftex]{graphicx}
\else
  \usepackage[dvips]{graphicx}
\fi
\usepackage{amsmath}

\usepackage{multicol, multirow, tabularx}
\usepackage[table]{xcolor}
\usepackage{xurl} %fixes the url breakings
\usepackage{csquotes}
\usepackage[normalem]{ulem}
\usepackage{tcolorbox}
\usepackage{caption}
\usepackage{graphicx}
\usepackage{makecell, subcaption}
\usepackage[hyphenbreaks]{breakurl}

\usepackage{url}
\usepackage{csquotes}
\usepackage{threeparttable}

\begin{document}
%\title{What Challenges Do Developers Face About Managing Secrets-Containing Software Artifacts?}

\title{A Comparative Study of Software Secrets Reporting by Secret Detection Tools}

\author{\IEEEauthorblockN{Setu Kumar Basak\IEEEauthorrefmark{1},
Jamison Cox\IEEEauthorrefmark{2}, Bradley Reaves\IEEEauthorrefmark{3} and
Laurie Williams\IEEEauthorrefmark{4}}
\IEEEauthorblockA{North Carolina State University, USA\\
Email: \IEEEauthorrefmark{1}sbasak4@ncsu.edu,
\IEEEauthorrefmark{2}jcox3@ncsu.edu,
\IEEEauthorrefmark{3}bgreaves@ncsu.edu,
\IEEEauthorrefmark{4}lawilli3@ncsu.edu}}

% \author{\IEEEauthorblockN{Anonymous Author(s)}
% }


%Telephone: (800) 555--1212, Fax: (888) 555--1212}


\maketitle

\begin{abstract}
\textit{Background:} According to GitGuardian’s monitoring of public GitHub repositories, secrets sprawl continued accelerating in 2022 by 67\% compared to 2021, exposing over 10 million secrets (API keys and other credentials). Though many open-source and proprietary secret detection tools are available, these tools output many false positives, making it difficult for developers to take action and teams to choose one tool out of many. To our knowledge, the secret detection tools are not yet compared and evaluated. \textit{Aims:} \textit{The goal of our study is to aid developers in choosing a secret detection tool to reduce the exposure of secrets through an empirical investigation of existing secret detection tools.} \textit{Method:} We present an evaluation of five open-source and four proprietary tools against a benchmark dataset. \textit{Results:} The top three tools based on precision are: GitHub Secret Scanner (75\%), Gitleaks (46\%), and Commercial X (25\%), and based on recall are: Gitleaks (88\%), SpectralOps (67\%) and TruffleHog (52\%). Our manual analysis of reported secrets reveals that false positives are due to employing generic regular expressions and ineffective entropy calculation. In contrast, false negatives are due to faulty regular expressions, skipping specific file types, and insufficient rulesets. \textit{Conclusions:} We recommend developers choose tools based on secret types present in their projects to prevent missing secrets. In addition, we recommend tool vendors update detection rules periodically and correctly employ secret verification mechanisms by collaborating with API vendors to improve accuracy. 
\end{abstract}


% \fi
\IEEEpeerreviewmaketitle

\section{Introduction} \label{Introduction}
%This demo file is intended to serve as a ``starter file''
%for IEEE conference papers produced under \LaTeX\ using
%IEEEtran.cls version 1.8b and later.
GitGuardian measured the exposure of secrets in GitHub repositories for the last three years and reported in March 2023 that secrets sprawl continued accelerating in 2022 by 67\% compared to 2021, exposing more than 10 million secrets~\cite{gitguardian-secret-sprawl}. In addition, they discovered that one out of 10 GitHub code authors exposed at least one secret in 2022. Secrets (such as API keys and access tokens) are indispensable for software as secrets are needed for third-party service integration, such as payment systems. However, developers leak secrets in plain text in the version control systems (VCS) and application packages~\cite{meli2019bad,android-leak}. In September 2022, an attacker took over Uber's internal tools and applications by leveraging hard-coded admin credentials in their PowerShell scripts~\cite{uber-breach}.

To prevent secrets from leaking in VCS, several open-source and proprietary tools such as Gitleaks and SpectralOps are available. However, these tools generate many false positives. Chess and McGraw~\cite{chessgarryalert} state that a high percentage of false positives may lead to 100 percent false negatives because people stop using the tool. This phenomenon is called \emph{alert fatigue}~\cite{alert-fatigue}. In addition, a tool will be unsound if it allows false negatives to escape to reduce false positives. As a result, developers face challenges in selecting secret detection tools. To our knowledge, no research has been conducted yet evaluating and comparing existing secret detection tools.

\textit{The goal of our study is to aid developers in choosing a secret detection tool to reduce the exposure of secrets through an empirical investigation of existing secret detection tools.}

In this study, we analyzed existing open-source and proprietary secret detection tools and provided answers to the following research questions:

\begin{itemize}
  \item \textbf{RQ1:} How do the secret detection tools perform in detecting secrets in terms of precision and recall?
  \item \textbf{RQ2:} What features are offered by the secret detection tools to aid in preventing secrets exposure?
\end{itemize}

We selected five open-source and four proprietary tools and compared the tools against a benchmark dataset of 818 repositories. We analyzed the tools report and evaluated how tools perform in detecting secrets. In addition, we analyzed the features offered by the tools in preventing the exposure of secrets and identified future research needs for secure software secret management. We have also made a dataset of the false positive secrets reported by the tools publicly-available for future researchers to aid in expediting research on the accuracy of the tools~\cite{fp-secretbench}. We summarize our contributions as follows:

\begin{itemize}
  \item A first comparative study of the existing open-source and proprietary secret detection tools and a qualitative analysis of the reports generated by the tools;
  \item A categorization of the features provided by the secret detection tools to aid in preventing secrets exposure; and
  \item A dataset of false positive secrets reported by the tools.
\end{itemize}

The rest of our paper is structured as follows: Section~\ref{BenchmarkDataset},~\ref{SecretDetectionTools} and~\ref{ComparingToolResult} introduce the benchmark dataset, selection process of tools, and the methodology to compare and evaluate the tools result, respectively. We discuss the findings and implications of our work in Section~\ref{Results} and~\ref{Discussion}. Section~\ref{Ethics} discusses the ethics, followed by the limitation of our paper. We discuss the related work in Section~\ref{RelatedWork} and conclude in Section~\ref{Conclusion}.

\section{Benchmark Dataset} \label{BenchmarkDataset}
To compare the secret detection tools, we selected \emph{SecretBench}~\cite{secretbench}, a publicly-available benchmark dataset of software secrets. We accessed the dataset using Google Cloud Storage (Bucket Name: \emph{secretbench})~\cite{google-cloud-storage} and Google BigQuery (Dataset ID: \emph{dev-range-332204.secretbench.secrets})~\cite{google-big-query}. A detailed description of the dataset is given below: 

\textbf{Repositories:} The dataset has been curated from the Google BigQuery Public Dataset of GitHub~\cite{github-big-query} using 761 regular expression patterns of different types of secrets. The dataset consists of 818 public GitHub repositories.

\textbf{Secrets:} The dataset consists of 97,479 labeled plain-text secrets (labeled as true and false) extracted from 818 repositories. The secrets were manually labeled by the two authors of SecretBench~\cite{secretbench}. Among the 97,479 candidate secrets, 15,084 are true secrets. In addition, among the true secrets, 4,457 are unique since the same secret can have multiple instances in a repository (multiple commits and files). 

\textbf{Categories}: The secrets of the dataset are categorized into eight categories. The number of total candidate secrets and true secrets of the eight categories are presented in Table~\ref{secret-categories}. The top three categories based on the number of true secrets are: ``Private Key", ``API Key and Secret" and ``Authentication Key and Token". The candidate secrets of the ``Other'' category are random strings and non-exploitable IDs such as GitHub commit IDs which are mostly false positives (99.29\%).

\newcolumntype{e}{>{\hsize=0.2\hsize}X}
\newcolumntype{s}{>{\hsize=0.6\hsize}X}
\newcolumntype{a}{>{\hsize=0.2\hsize}X}
\begin{table} [!htb]
%\small
\footnotesize
\caption{The eight categories of secrets in SecretBench.}
\label{secret-categories}
\begin{tabularx}{\columnwidth} {|s | e |e|}
 \hline
 \multicolumn{1}{|l|}{\textbf{Category}} &
 \multicolumn{1}{l|}{\textbf{True Secrets}} &
 \multicolumn{1}{l|}{\textbf{Total Secrets}}\\
 \hline \hline
 Private Key & 5,789 & 8,584 \\ \hline
 API Key and Secret & 4,529 & 5,162 \\ \hline
 Authentication Key and Token & 3,569 & 5,833 \\ \hline
 Other & 524 & 66,690 \\ \hline
 Generic Secret &  334 & 439 \\ \hline
 Database and Server URL & 162 & 9,970 \\ \hline
 Password & 150 & 705 \\ \hline
 Username & 27 & 96 \\ \hline
\end{tabularx}
\end{table}

\textbf{Programming Languages:} The dataset repositories comprised source codes of 49 programming languages. The top five programming languages based on the number of repositories are Shell (459), JavaScript (414), Python (312), Java (180), and Ruby (172). The number in the parenthesis denotes the number of repositories that used the specific language.

\textbf{File Types:} The dataset consists of 311 file types in which secrets have been found. All the 311 file types and the number of true secrets present in these file types can be found in the GitHub repository of SecretBench~\cite{secretbench-file-types-secrets}. The top five file types based on the number of true secrets are presented in Table~\ref{secret-file-types}. 

\newcolumntype{e}{>{\hsize=0.1\hsize}X}
\newcolumntype{s}{>{\hsize=0.7\hsize}X}
\newcolumntype{a}{>{\hsize=0.2\hsize}X}
\begin{table} [!htb]
%\small
\footnotesize
\caption{SecretBench's top five file types on true secrets.}
\label{secret-file-types}
\begin{tabularx}{\columnwidth} {|e | s |e|}
 \hline
 \multicolumn{1}{|l|}{\textbf{File Type}} &
 \multicolumn{1}{l|}{\textbf{Description}} &
 \multicolumn{1}{l|}{\textbf{True Secrets}}\\
 \hline \hline
 txt & Text File & 2,935 \\ \hline
 toml & Configuration File & 1,985 \\ \hline
 js & Javascript file & 1,583 \\ \hline
 html & Hypertext Markup Language File & 1,337 \\ \hline
 pem &  Privacy Enhanced Mail Format File & 813 \\ \hline
\end{tabularx}
\end{table}

\textbf{Secrets Metadata:} The dataset provides secrets metadata, such as repository name, file path, commit id and start line of where the secrets are matched. We used the metadata to compare the tool-reported secrets, as discussed in Section~\ref{ComparingToolResult}.

\section{Secret Detection Tools} \label{SecretDetectionTools}
In this section, we explain the selection process of secret detection tools; provide a brief description of each tool; how we installed each tool; and how we scanned the benchmark repositories using each tool.

\subsection{Selection of Secret Detection Tools}\label{selectionoftools}
To find the existing open-source and proprietary secret detection tools, we searched both the web and academic literature. We constructed a set of the following search strings: \emph{(secret OR credential OR password) AND (detection OR scanning OR digger) AND (tool OR utility)}. For web search, we used the Google Search Engine and selected the top 100 results for each search string according to the Google Search Engine's Page Rank algorithm. The stopping criteria of 100 for each search string has been set based on the guideline of grey literature search in prior works~\cite{GAROUSI2019101}. Similarly, for academic literature search, we searched the top five scholar databases recommended in the computing science domain~\cite{acm, springer, ieeexplore, dblp, sciencedirect}. We identified 20 tools from the search result and applied the following selection criteria to choose the secret detection tools for our study.

\begin{enumerate}
    \item \textbf{Accessible:} The tool can be installed into a local system or accessed via subscription from the tool vendors. 
    \item \textbf{Scans Git Repositories:} The tool can scan Git repositories since our dataset contains Git repositories.
     \item \textbf{Active:} The tool's repository has shown activity for the last two years. We checked the last commit date in the repository of the open-source tools.
    \item \textbf{Flags Secrets:} The tool flags individual secrets instead of flagging only secret-containing suspicious file names.
    \item \textbf{Reports Plain Text Secret:} The tool reports secrets in plain text as we must compare the secrets with our benchmark dataset.
\end{enumerate}

Based on the above selection criteria, we excluded 11 tools.  After each tool, we provide in parenthesis the criteria we used to exclude a tool using the enumerated criteria listed above: Credential-Digger~\cite{credential-digger} (1), Credscan~\cite{credscan} (1), Cycode~\cite{cycode} (1), detect-secrets~\cite{detect-secrets} (5), git-all-secrets~\cite{git-all-secrets} (3), git-hound~\cite{git-hound} (5), gitrob~\cite{gitrob} (3), Gittyleaks~\cite{gittyleaks} (3), repo-security-scanner~\cite{repo-security-scanner} (4), SecretHunter~\cite{secrethunter} (1) and Saha et al. Tool~\cite{saha2020secrets} (1). %The number in parenthesis denotes the selection criteria for which the tools have been excluded. 
Ultimately, we selected 9 secret detection tools, of which 5 tools are open-source and 4 tools are proprietary.

\subsection{Tools Description}

For the selected secret detection tools, we provide a) a brief description of the tool, b) how we installed the tool, and c) the scanning technique employed for finding secrets in benchmark repositories. Since each tool provides configuration options for detecting secrets, we installed and ran the tools with recommended configurations by contacting the tool vendors or by obtaining suggested configurations in the product documentation to get higher accuracy.

\textbf{git-secrets:} git-secrets~\cite{git-secrets} developed by AWS-Labs~\cite{aws-labs} is an open-source tool. We installed Version \texttt{1.3.0} of the tool using HomeBrew. In addition, as a pre-requisite to scan for secrets in the repositories, we installed two git hooks (\texttt{git secrets --install} and \texttt{git secrets --register-aws}) separately for each repository. We used the \texttt{--scan-history} flag (\texttt{git secrets --scan-history \&> report.txt}) to scan the entire Git history and outputted the secrets in a text file.

\textbf{Gitleaks:} Gitleaks~\cite{gitleaks} is an open-source tool written in Go. We installed Version \texttt{8.2.7} of the tool using HomeBrew and scanned the repositories using the \texttt{detect} command (\texttt{gitleaks detect -v --source=repo\_dir --report-path=report.json}). The verbose flag (\texttt{-v}) has been used to retrieve metadata information of the matched secret, and we extracted the secrets in JSON files.

\textbf{Repo-supervisor:} Repo-supervisor~\cite{repo-supervisor} is an open-source tool written in JavaScript. We downloaded the binary release (Version \texttt{3.2.0}) and installed Node Package Manager (NPM) dependencies (\texttt{npm ci \&\& npm run build}). The tool operates in two separate modes. The first mode allows to scan GitHub pull requests through webhooks, and the second mode works from the command line, where it scans local repository directories. We performed the latter by executing the \texttt{cli.js} file (\texttt{JSON\_OUTPUT=1 node ./dist/cli.js repo\_dir}) and extracted the output in JSON file.

\textbf{TruffleHog:} TruffleHog~\cite{trufflehog} is an open-source tool developed by Truffle Security~\cite{trufflesecurity} and written in Go. We installed Version \texttt{3.18.0} of the tool using HomeBrew. We scanned the repositories with \texttt{--regex} and \texttt{--entropy} flags enabled (\texttt{trufflehog git --regex --entropy file://repo\_dir}) and downloaded the JSON report.

\textbf{Whispers:} Whispers~\cite{whispers} is an open-source tool written in Python. The tool supports different formats for structured text parsing, such as YAML and XML. The tool parses the source code in key-value pairs, where the key is the field name and the value is the potentially hard-coded secret assigned to the given key. We installed Version \texttt{2.1.5} of the tool using \texttt{pip3}. To scan the repositories, we executed the \texttt{whispers repo\_dir > report.json} command and extracted the output in JSON files.  

\textbf{Commercial X:} Since the proprietary tool vendor would not allow their identity to be disclosed in the paper, we refer to them as ``Commercial X''. In addition to scanning GitHub repositories, the tool can find secrets in images and non-searchable PDFs. The tool can be integrated with Slack, JIRA, and Google Drive to find any secrets exposure. We contacted their team and provided the snapshot of 818 repositories of our benchmark. They ran their tool on those repositories and provided us with the scan report. We parsed the scan report and outputted the secrets with the metadata in a CSV file.

\textbf{ggshield:} ggshield~\cite{ggshield} has been developed by GitGuardian~\cite{GitGuardian}. We installed the tool (Version \texttt{1.14.3}) using HomeBrew. Though the tool is open-sourced in GitHub, the tool requires an API key for scanning a repository since ggshield internally uses GitGuardian's public API~\cite{gitguardian-api} through py-gitguardian~\cite{py-gitguardian} client to scan and detect secrets. We contacted GitGuardian to get an API key (API Quota Limit: 8 Million) and set the key in the local environment variable to scan all the benchmark repositories. We executed the \texttt{scan repo} command (\texttt{ggshield secret scan repo repo\_dir --show-secrets --json -v -o report.json}) for searching secrets in each repository. The \texttt{--show-secrets} flag has been used to extract the secrets in non-redacted form, and the found secrets are outputted in a JSON file.

\textbf{Github Secret Scanner:} GitHub has an integrated secret scanner~\cite{github-secret-scanner} to scan for secrets in the repositories. The ``Secret Scanner" settings can be enabled from the ``Code security and analysis'' option in GitHub. To scan the repositories of the benchmark dataset, we forked each repository into the first author's GitHub account. We enabled the ``Secret Scanner" settings for each repository. As soon as we enabled the settings, the scanner was triggered and displayed the detected secrets under the ``Security/Secret scanning alerts" tab of the specific repository. We wrote a Python script to extract each repository's secrets in a CSV file using GitHub Rest API~\cite{github-rest-api}.

\textbf{SpectralOps:} SpectralOps~\cite{spectralops} is a proprietary tool. To scan repositories in a local environment, we created a Spectral account and contacted the Spectral support team to gain access for seven days. We received a Spectral Data Source Name (DSN) key and saved it in the local environment. The tool provides three scanning modes: ``Developer", ``Security" and ``Audit" based on different precision and recall rates. The Spectral team recommended using the ``Security" mode for better precision and recall. We ran the scan command (\texttt{spectral scan --all --forensic --ok --show-match --include-tags base,audit --with-branches --json  report.json}) and outputted secrets in JSON files. The base and audit tags are used for ``Security" scan mode, and \texttt{--forensic} flag retrieves the secret's metadata.

\subsection{Machine Configuration}

We installed eight tools in two Mac instances except for the GitHub Secret Scanner and Commercial X. The configuration of the instances are as follows: Instance 1 (OS: Monterey version 12.3.1, RAM: 64 GB, Persistent Disk: 1 TB) and Instance 2 (OS: Monterey version 12.6.2, RAM: 32 GB, Persistent Disk: 1 TB). We used two Mac instances to speed up the scanning process since the benchmark dataset contains large repositories with a large commit count. After scanning with each tool, we wrote Python scripts to extract the secret with additional metadata from the JSON and text files and outputted in CSV files. The extracted results are used for analysis and comparison, as discussed in Section~\ref{ComparingToolResult}.

\section{Analyzing Tool Results} \label{ComparingToolResult}
In this section, we explain the secret and tool metadata we analyzed and how we filtered and compared the tool results to answer our research questions.

\subsection{Secret Metadata}\label{SecretsMetadata}

Below, we discuss the metadata information related to secrets we processed to answer our research questions.

\textbf{Commit ID:} A commit id in Git is a unique SHA-1 hash created whenever a new commit is recorded. The commit id helps to identify the exact commit reference where the secrets have been found during comparison.

\textbf{File Path:} The file path is the file's location in the repository where the secret has been found. We normalized the file path as it contained either the computer root folder location where the tool has been installed or the repository directory. For example: Repo-supervisor outputs the file path as ``\texttt{<Repo\_dir>\slash conf\slash file.py}'' while Spectralops outputs as ``\texttt{/Users\slash<User\_name>\slash<Repo\_dir>\slash conf\slash file.py}''. We extracted the file path as ``\texttt{conf\slash file.py}'' for comparison.   

\textbf{Line Number:} The line number denotes the line in the file where the secret has been matched, which helps to identify if the same secret is present in multiple places of the same file. 

\textbf{Plain Text Secret:} The plain text secret is the tool-reported hard-coded secret in the source code. However, some tools report secrets along with the source code context. For example, git-secrets outputs the function or variable declaration where the secret is used (\texttt{bitly\_token <- bitly\_auth(key = "xxxxxx")}). The ``\texttt{xxxxxx}'' is the secret where \texttt{bitly\_auth} and \texttt{bitly\_token} are the function and variable name, respectively. As a result, matching reported secrets with the benchmark through automation is difficult. In addition, manual inspection is impractical due to the large number of reported secrets by the tools. However, we observed patterns such as ``\texttt{key=}", ``\texttt{token=}" and ``\texttt{id:}" in the reported secret text. We removed non-alphanumeric characters, such as brackets and space, from the string and extracted the secret by only taking the string part after the pattern. We used these normalized secrets for comparison.

\textbf{Alert Count:} The alert count is the total number of alerts reported by each tool which indicates the amount of audit effort required by the practitioners. Tools such as SpectralOps and ggshield provide the number of alerts in the respective reports. For tools that do not provide the number of alerts in the report, we calculated the total number of alerts using a Python script by iterating through each report.

\subsection{Filter and Compare Tool Alerts}\label{FilterCompareToolAlerts}

We observe that tools provide non-secret alerts, such as alerts for suspicious files and dangerous functions. For example, Whispers flags suspicious files, such as \texttt{database.sql} file, and dangerous functions, such as \texttt{exec} and \texttt{eval}. In the output, the tool provides a rule identifier for different types of alerts, such as \texttt{secret} and \texttt{api-key} for secrets; \texttt{file-known} for suspicious files; and \texttt{system} for dangerous functions. We filtered the non-secret alerts using the rule identifiers. We also filtered secrets committed after November 25, 2022, since the benchmark dataset contains secrets introduced before that date. For example, the GitHub secret scanner scans the repository's latest snapshot (February 25, 2023) since the tool can not scan a local repository. We retrieved the commit date of each commit using GitHub Rest API~\cite{github-rest-api}. We filtered any secrets introduced after November 25, 2022, for a fair comparison of the tools with the benchmark.

Next, we compare the secret of each repository reported by the tool with the secrets of the same repository in the benchmark. We mark the secret reported by the tool as true positive (TP) if the secret is matched. Otherwise, we mark the secret as a false positive (FP). However, we are unable to match different types of secrets with exact string comparison for all the tools though we normalized the secrets. Below, we discuss the different scenarios of the secret match and how we calculated the match for each.

\textbf{Jaro-Winkler Similarity}: After normalizing the secrets for source code context, we observe that additional source code as a suffix can be present. For example, git-secrets outputs secrets with additional source code context (\texttt{"analytics\_configuration": \{key: "xxxxxxxxxxxxx", type: "Traffic"\}}). The secret is ``\texttt{xxxxxxxxxxxxx}'' and after normalizing, we got ``\texttt{xxxxxxxxxxxxxtypeTraffic}'' where the string part ``typeTraffic" is not part of the secret. As a result, we cannot perform an exact match of the secret with the benchmark. To address this scenario, we used Jaro-Winkler Similarity~\cite{winkler1990string} for string comparison, a variant of the Jaro Distance metric~\cite{jaro1989advances}. The Jaro–Winkler similarity employs a prefix scale that rewards strings that match from the beginning with high scores~\cite{winkler1990string}. The Jaro–Winkler algorithm provides a similarity score between [0,1] where 0 represents two entirely dissimilar strings and 1 represents identical strings. We used the \texttt{jaro\_winkler\_similarity} function of \texttt{jellyfish}~\cite{jellyfish-python} package in Python to calculate the similarity. We found the similarity score of ``\texttt{xxxxxxxxxxxxx}'' and  ``\texttt{xxxxxxxxxxxxxtypeTraffic}'' is 0.82. We termed two secrets a match if the similarity score equals or exceeds 0.7. We set the cut-off similarity score of 0.7 by randomly sampling secrets and observing the score with the benchmark.

\begin{figure}
    \includegraphics[width=\columnwidth]{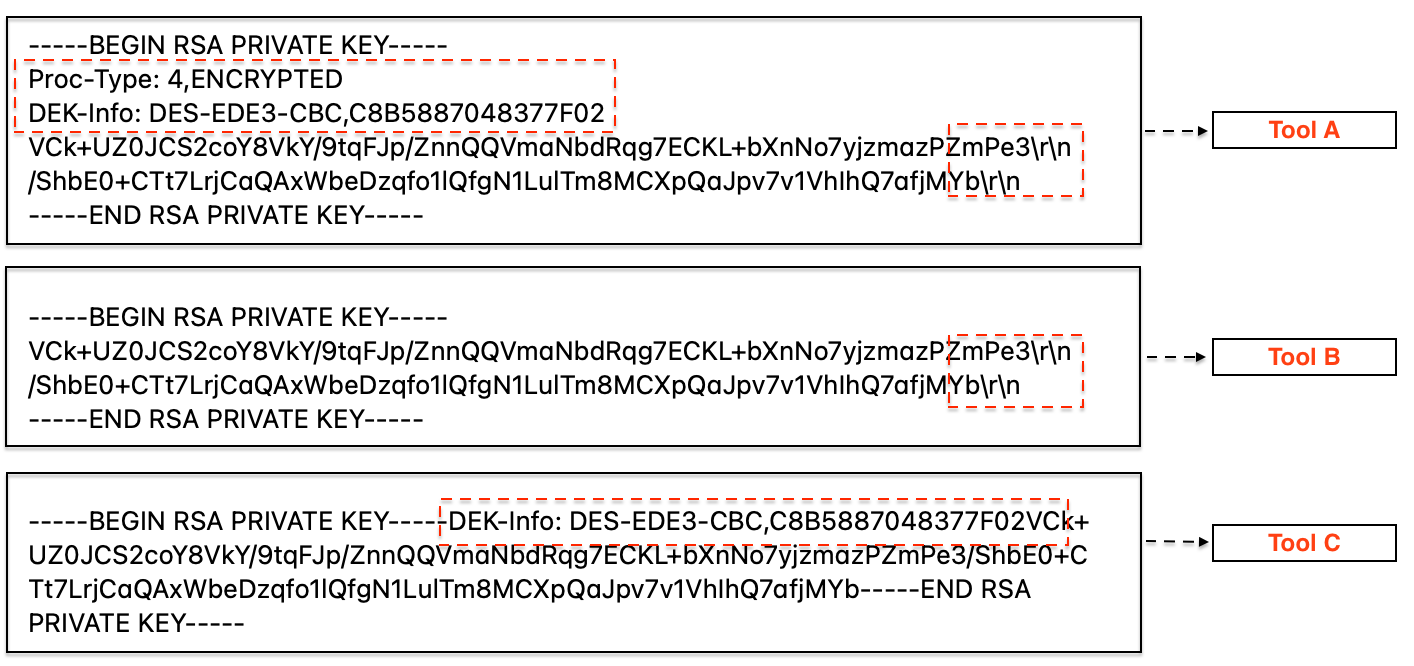}
    \caption{Different outputs of the same secret by three tools.}
    \label{fig:multi-line-secret}  
\end{figure}

\textbf{Gestalt Pattern Match:} We observe that a secret can contain additional context in the middle, especially for multi-line secrets. For example, private keys are generally present as multi-line in the source code. Tools output these private keys differently, making it difficult to perform an exact match with the benchmark. Figure~\ref{fig:multi-line-secret} shows three different outputs of the same secret. Tool A outputs the ``Proc-Type" and ``DEK-Info" properties along with carriage return (``\texttt{\textbackslash r}") and line feed (``\texttt{\textbackslash n}"), which is the same as the benchmark. However, Tool B excludes the ``Proc-Type" and ``DEK-Info" properties in the output, and Tool C includes the properties but outputs the secrets in a single-line instead of a multi-line without ``\texttt{\textbackslash r\textbackslash n}". To address this scenario, we used the Gestalt pattern matching algorithm~\cite{black2004ratcliff} after removing non-alphanumeric characters from the secret and making the secret single-line. The algorithm calculates the similarity score by finding the longest common substring and then recursively finding the number of matching characters in the non-matching regions on both sides of the longest common substring~\cite{black2004ratcliff}. As a result, we could match a secret even if the secret does not contain the middle context (the properties of the private key). We used the \texttt{SequenceMatcher} function of \texttt{difflib}~\cite{sequencematcher} package in Python to calculate the Gestalt similarity score. We termed two secrets a match if the similarity score equals or exceeds 0.6. Similar to the Jaro-Winkler similarity, we set the cut-off similarity score of 0.6 by randomly sampling secrets and observing the score with benchmark secrets.

We marked a secret reported by a tool as TP if the secret equals or exceeds the cut-off similarity score of either the Jaro-Winkler or Gestalt algorithm. To check whether the combination of algorithms correctly matches tool-reported secrets with benchmark and label automatically, we randomly selected 100 unique reported secrets from each tool and manually inspected the label calculated by the algorithms. The combination of both algorithms correctly labeled 97\% of the secrets.  

\textbf{Recall Cases:} We observe that the same secret can be present in multiple commits, multiple files, and different lines of the same file of a repository. As a result, finding and removing all instances of a secret from the source code is necessary. However, every tool does not provide all the metadata related to secrets, such as the commit id, file path, and line number, as shown in Table~\ref{tools-feature}. As a result, we calculated the recall of each tool in two cases to have a fair comparison. Case 1 of recall denotes when the secrets of the benchmark are found exactly in the same commit, file, and line number of the tools report, and Case 2 denotes that the secrets of the benchmark are found at least in the repository, irrespective of the metadata. For Case 1, we matched each tool's reported secrets with all the benchmark secrets for a repository. If a secret of the benchmark matches the tool-reported secret but does not match the metadata, then we mark the secret as a false negative (``FN''). However, for Case 2, we matched the unique secrets of the benchmark for a repository with the reported secrets of the tools. If a secret of the benchmark matches the tool-reported secret but does not match the metadata, we still mark the secret as true positive (``TP'') since the secret is at least found in the repository. We could not calculate Case 1 for Repo-supervisor and SpectralOps as these tools do not provide either commit id or line number, thus calculating F1-score using precision and Case 2 of recall.

\subsection{Tool Metric}

Below we discuss the tool metric we calculated to answer our research questions.

\textbf{Scan Time:} Scan time helps to understand how quickly secrets will be identified to remediate any secrets exposure. Running each tool multiple times on all 818 benchmark repositories is impractical since scanning takes a long time. Hence, we calculated the scan time on a sample set of repositories of our benchmark to calculate the efficiency of the tools. First, we curated the sample set of 15 repositories as follows: 

\begin{itemize}
    \item \textbf{Repository Size:} The largest, smallest and median size of a repository in the benchmark is 5,658.22 MB, 0.04 MB, and 37.42 MB, respectively. We selected a random sample of 6 repositories based on the repository size: 4 repositories with repository sizes greater than the median and 2 repositories less than the median.
    \item \textbf{Commit Count:} Since a repository of a larger size can have a low number of commit counts, and vice-versa, we also included repositories in the sample set based on the commit count. The benchmark repository's highest, lowest and median commit count is 425,699, 22, and 1,200, respectively. We selected a random sample of 6 repositories based on the commit count: 4 repositories with a commit count greater than the median and 2 repositories less than the median.
    \item \textbf{Programming Language:} The sample set should have at least 1 repository for each of the top 5 programming languages of the benchmark (see, Section~\ref{BenchmarkDataset}). We randomly selected 3 additional repositories since 2 languages were already present in the above-selected 12 repositories.
\end{itemize}

Next, we ran each tool 5 times on each of the 15 repositories, calculated the total scan time using the \texttt{time}~\cite{time-python} package of Python, and calculated the average scan time.

\textbf{Popularity:} Since the open-source tools publish their source code in a public repository, we can measure the tool's popularity among the developers. Developers can \emph{fork} the open-source tools repository in GitHub. The fork count of a repository indicates a higher chance of attracting potential contributors to the project. Developers can also \emph{star} a repository when they want to appreciate the project and \emph{watch} when they want to be notified of all the activities (bug fixes, new features) of the project. We used each open-source tools repository's fork, star, and watch count as a proxy to calculate the tool's popularity instead of considering a single metric. Previous studies~\cite{github-popularity, BORGES2018112} have also used these metrics to calculate the popularity of a repository. To verify the rank correlation among fork, star, and watch count, we calculated the Spearman's rho ($\rho$)~\cite{spearmanrho} using Kaggle's GitHub repository dataset~\cite{kaggle-github}. We observed a significant correlation between star and fork ($\rho$ = 0.71), watch and fork ($\rho$ = 0.60), and watch and star ($\rho$ = 0.55) counts. To calculate the popularity score for each tool, we normalized the fork, star, and watch counts using min-max normalization~\cite{featurescale} and calculated the average of the counts.

\section{Results} \label{Results}
In this section, we discuss our findings and answer our research questions.

\subsection{RQ1: How do the secret detection tools perform in detecting secrets, in terms of precision and recall?}\label{RQ1}

Below we discuss a) the precision, recall, and F1-score of each tool; b) the overlap of secrets reporting by the tools; c) a comparison of the scan time and popularity of the tools; and d) an analysis of the false positives and false negatives reported by the tools.

\textbf{\underline{Precision, Recall and F1-score:}} Table~\ref{tool-accuracy} presents the precision, recall and F1-score of each tool. The column ``Precision (Total Alerts, TP)'' denotes the precision of each tool in detecting secrets. The numbers in parenthesis denote the total number of alerts reported by the tool and the count of true positives detected by the tool, respectively. The columns ``Recall - Case X (TP, FN)'' present the recall of each tool, where X denotes the two cases as discussed in Section~\ref{FilterCompareToolAlerts}. The numbers in parentheses denote the number of true positives and false negatives found by the specific tool, respectively. Low precision indicates more false positives causing the tool to be unusable and low recall indicates more false negatives causing a missed opportunity to be alerted of a secret. The column ``F1 Score'' denotes the F1-score of each tool, the harmonic mean of precision and recall (Case 2) as discussed in Section~\ref{FilterCompareToolAlerts}. Below, we discuss our observations related to precision, recall, and F1-score.

\begin{table} [!htb]
\footnotesize
\centering
\begin{center}
\caption{Precision, Recall, F1-Score, Scan Time (ST), and Popularity Score (PS) of each tool.}
\label{tool-accuracy}
    \includegraphics[width=\columnwidth]{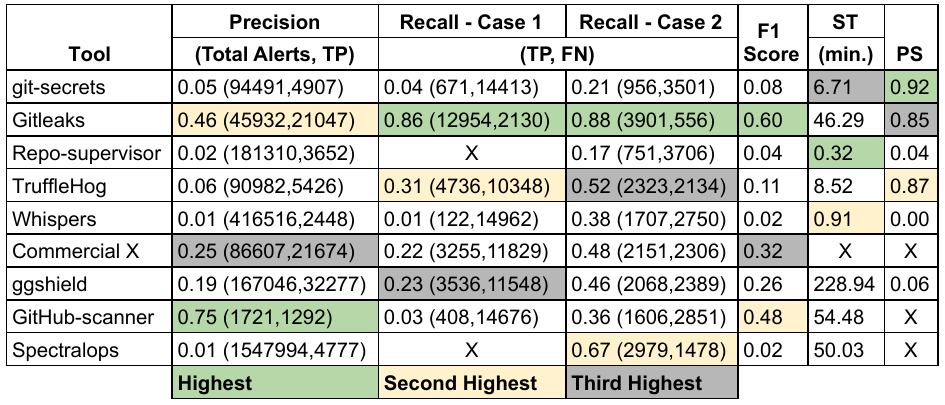}  
\end{center}

\end{table}

\begin{itemize}
    \item We observe that based on the precision, the top three tools are GitHub Secret Scanner (75\%), Gitleaks (46\%), and Commercial X (25\%), respectively. Among the nine tools, five tools have a precision score of less than 7\%.
    \item Based on recall, we observe that Gitleaks is the top tool in both cases (Case 1: 86\% and Case 2: 88\%) and the second-best based on precision. In addition, TruffleHog has the second-best recall in Case 1 (31\%) and third-best in Case 2 (52\%) though the precision is only 6\%.
    \item We observe that based on F1-score, the top three tools are Gitleaks (60\%), GitHub Secret Scanner (48\%), and Commercial X (32\%).
    \item Though GitHub Secret Scanner is the top tool based on precision, the recall score is low (6\%), indicating the tool misses many secrets. In contrast, SpectralsOps is the third-best based on recall (68\%), with a precision score of only 1\%. Thus, our findings indicate that no current tool has the coveted high precision and high recall scores.
    \item Recent research~\cite {9794113,saha2020secrets} utilizes machine learning (ML) to reduce false positives. However, Commercial X and SpectralOps, which employ ML to detect secrets, have lower precision scores 25\% and 1\%, respectively.
\end{itemize} 

\begin{table*} [!htb]
\footnotesize
\centering
\caption{Recall of each tool for eight secrets categories.}
\label{recall-secret-categories}
    \includegraphics[width=\textwidth]{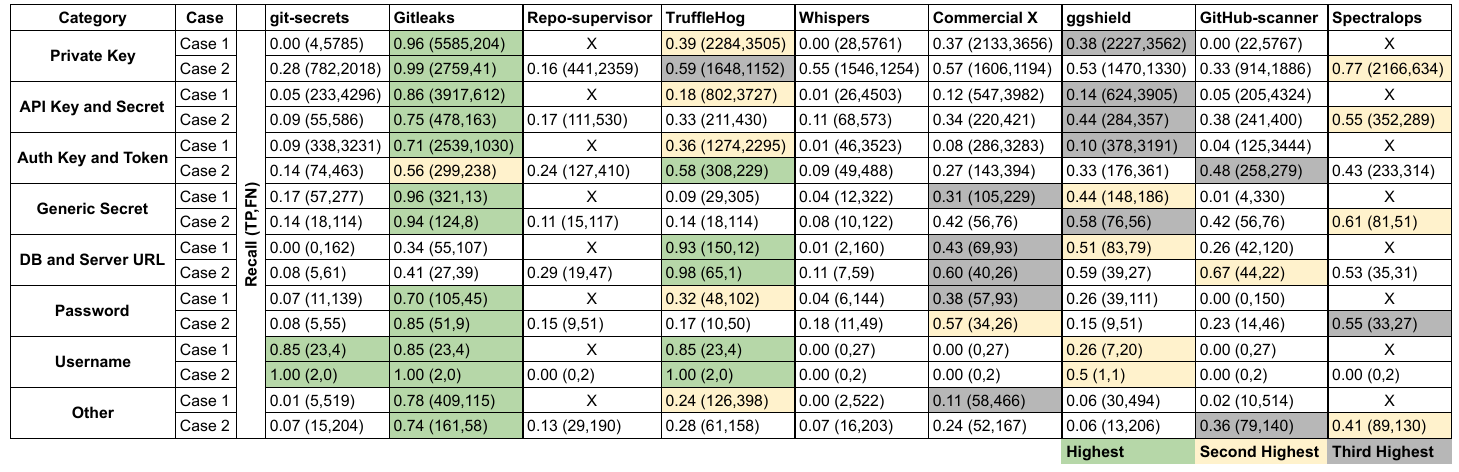}
\end{table*}

Since the secrets of our benchmark dataset are categorized into eight categories, such as ``Private Key'' and ``API Key and Secret'', we calculated the recall score per category for each tool. As a result, we identified which tool performs best in which category of secrets to aid developers in choosing tools based on the category of secrets present in their code. Table~\ref{recall-secret-categories} presents the recall score of each tool for the eight categories in two cases (Case 1 and Case 2). The numbers in parentheses denote the number of true positives and false negatives found by the tool for a specific category. We observed that Gitleaks and TruffleHog are the top two tools in most categories. However, SpectralOps has the second-best recall score for categories such as ``Private Key'' (Case 2) and ``Generic Secret'' (Case 2), whereas ggshield has the second-best recall for ``Username'' (Both cases). In addition, SpectralOps has the second-best recall score for categories such as ``API Key and Secret'' (Case 2), whereas GitHub Secret Scanner has the second-best recall for ``Database and Server URL'' (Case 2).

\begin{figure}
    \includegraphics[width=\columnwidth]{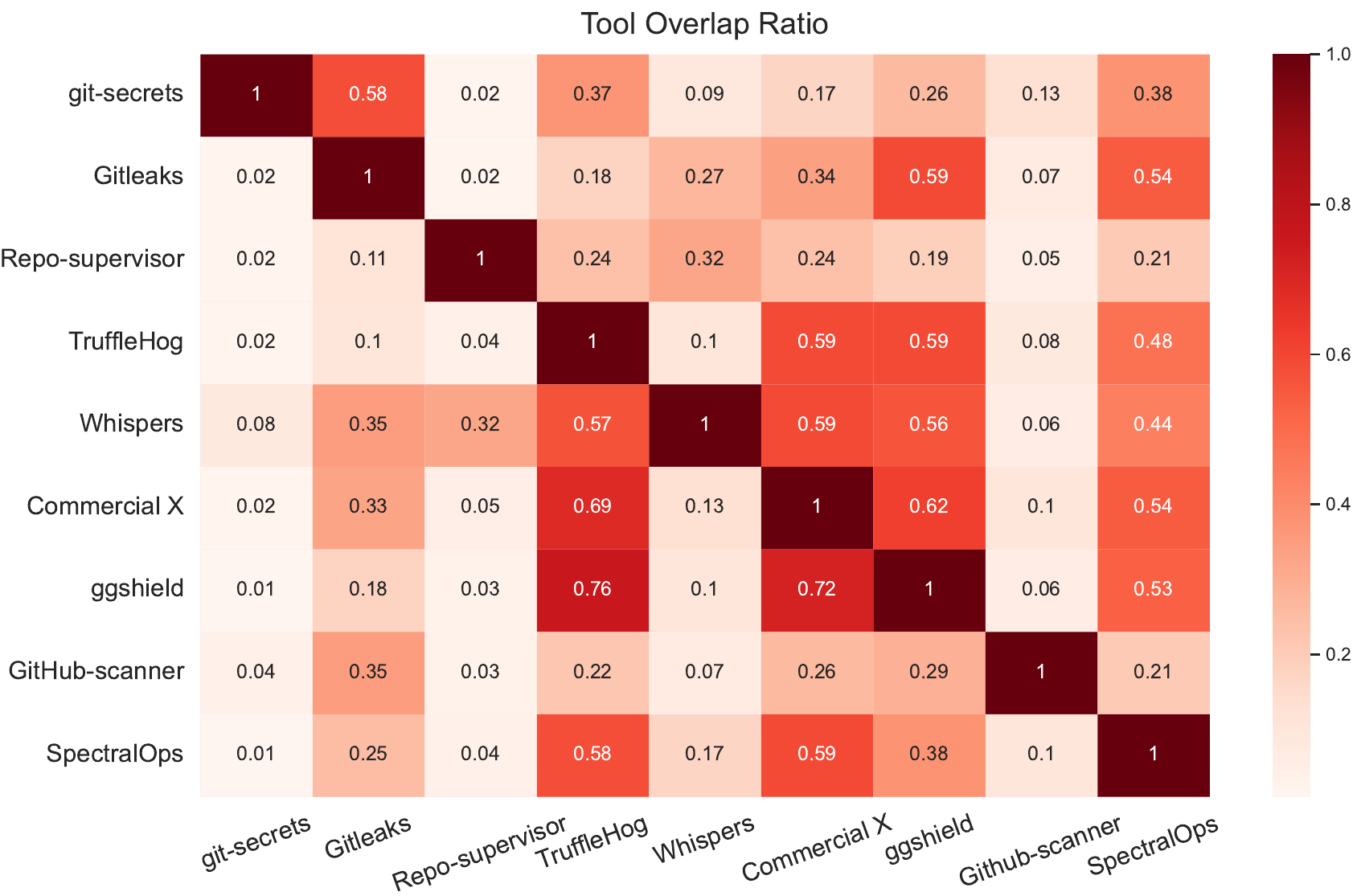}
    \caption{Overlap ratio of secrets reported by each tool.}
    \label{fig:tool-overlap}  
\end{figure}

\begin{figure}
    \centering
    \begin{subfigure}[t]{0.45\columnwidth}
        \centering
        \includegraphics[height=1.3in]{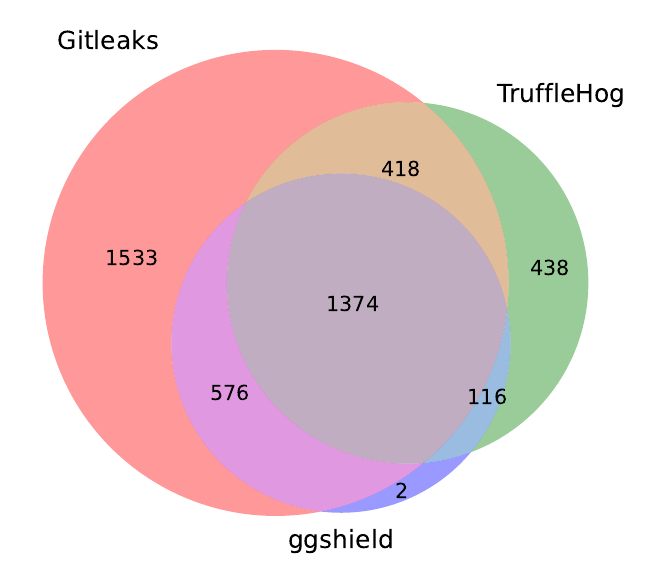}
        \caption{}
        \label{fig:tool-overlap-venn-a}
        %\caption{Overlap of secrets among Gitleaks, TruffleHog, and ggshield.}
    \end{subfigure}%
    %~ 
    \begin{subfigure}[t]{0.45\columnwidth}
        \centering
        \includegraphics[height=1.3in]{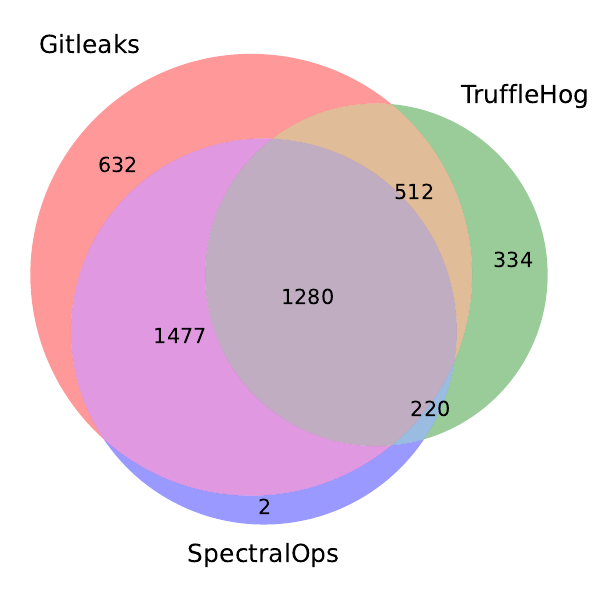}
        \caption{}
        \label{fig:tool-overlap-venn-b}
        %\caption{Overlap of secrets among Gitleaks, TruffleHog, and ggshield.}
    \end{subfigure}
    \caption{Venn diagram for overlap of unique true positive secrets among top three tools based on recall. Subfigure (\emph{a}) depicts the overlap of Gitleaks, TruffleHog, and ggshield. Subfigure (\emph{b}) depicts the overlap of Gitleaks, SpectralOps, and TruffleHog.}
    \label{fig:tool-overlap-venn}
\end{figure}

\textbf{\underline{Tool Overlap:}} We measured how much unique true positive (TP) secrets one tool reported overlap with another to identify which tools output similar secrets. The heatmap of Figure~\ref{fig:tool-overlap} depicts the overlap ratio between each pair of tools. For a pair of tools (A, B), the heatmap shows how many unique TP secrets reported by tool A are also reported by tool B. For example, 76\% of the unique TP secrets reported by ggshield are also reported by TruffleHog. However, only 18\% of the unique TP secrets reported by ggshield are reported by Gitleaks. The Venn diagrams in Figure~\ref{fig:tool-overlap-venn} show the non-overlap unique TP secrets among Gitleaks, TruffleHog, and ggshield (\emph{Top three tools based on recall (Case 1)}) and among Gitleaks, SpectralOps, and TruffleHog (\emph{Top three tools based on recall (Case 2)}). Figure~\ref{fig:tool-overlap-venn-a} shows that Gitleaks and TruffleHog outputs 1533 and 438 non-overlap unique TP secrets, respectively. Similarly, as shown in Figure~\ref{fig:tool-overlap-venn-b}, we observed that Gitleaks and TruffleHog outputs 632 and 334 non-overlap unique TP secrets, respectively. As a result, our findings substantiate the necessity of not relying on a single tool to identify all the secrets present in a repository.

\textbf{\underline{Scan Time:}} The column ``ST'' of Table~\ref{tool-accuracy} shows the time taken by each tool in minutes to scan the sample set of repositories. We could not calculate the scan time of Commercial X as the tool vendor has conducted the scanning, and the report does not contain any scan time. The top three tools based on scan time are Repo-supervisor, Whispers, and git-secrets, which took 0.32, 0.91, and 6.71 minutes, respectively. However, these tools have relatively low precision and recall scores indicating that tools did not scan the source code thoroughly. In contrast, the top two tools based on precision - GitHub Secret Scanner and Gitleaks took 54.48 and 46.29 minutes, respectively. However, we observe that tools having higher scan times do not always yield high precision and recall scores. For example, ggshield took the highest amount of time (4.8 hours) among all the tools, but the precision and recall were relatively low. We identified that Gitleaks, GitHub Secret Scanner, and SpectralOps showed a balance between scanning time and either high precision or recall. 

\textbf{\underline{Tool Popularity:}} The column ``PS'' of Table~\ref{tool-accuracy} presents the popularity score of each tool. We could only calculate the popularity score of the five open-source tools and one proprietary tool, ggshield. The source code of ggshield is open-sourced in GitHub, except for their proprietary scanning API implementations. Based on the PS score, the top three tools are git-secrets (0.92), Gitleaks (0.87), and TruffleHog (0.85), respectively. Though git-secrets is the most popular among the developers, the precision and recall are relatively low. In contrast, Gitleaks and TruffleHog are popular among developers having relatively high precision or recall scores.

\textbf{\underline{Analysis of False Positives:}} Since we observed a high false positive rate by the tools, we inspected a random sample of 50 false positives from each tool to identify the types of false positive secrets. Below, we discuss our observations related to the false positive secrets and the detection rules triggering the false positives.

\textit{\underline{1. Generic Regular Expressions (regex):}} Tools use generic regex to detect secrets that trigger false positives. Below we discuss the generic regex for different types of secrets.

\textbf{1.1 API Keys and Tokens:} Tools, such as Whispers, employ generic regex (\texttt{.*[A-Za-z0-9\-\_]+(key|token)\$}) for finding API keys and tokens. The regex treats any string having a ``key'' or ``token'' at the end as an API key or token. As a result, placeholder API keys or tokens such as ``\texttt{testkey}'' and ``\texttt{sampletoken}'' are output as secrets. However, tools such as Gitleaks and GitHub Secret Scanner identify API keys and tokens by applying regex for specific API keys and tokens. For example, the regex employed for the Stripe API key is \texttt{(?i)(sk|pk)\_(test|live)\_[0-9a-z]\{10,32\}}. However, the regex matches ``\texttt{sk\_live\_111111111111}'', a dummy API key, and outputs as a Stripe API key.

\textbf{1.2 Password:} To detect passwords, generic regex such as \texttt{(passwords?|passwd|pass|pwd)\_?[0-9]*\$} is used. As a result, strings such as ``\texttt{testpassword}'' or a UNIX command (``\texttt{pwd}'') are detected as passwords. 

\textbf{1.3 Cryptographic Key:} According to a study by Meli et al.~\cite{meli2019bad}, cryptographic keys are the most exposed secrets in the source code. However, tools employ generic regex (\texttt{.*[-]\{3,\}BEGIN (RSA|DSA|EC|OPENSSH)? ?(PRIVATE)? KEY[-]\{3,\}.*}) to identify cryptographic keys, thus reporting false positives. For example, a template string such as ``\texttt{---BEGIN RSA KEY---}'' with no following RSA key characters matches as a secret.

\textit{\underline{2. Ineffective Entropy Calculation:}} We observe that tools employ Shannon entropy~\cite{shannon1948mathematical} to identify possible secrets. Though the core Shannon entropy algorithm is correct, differentiating secrets from false positives is not always effective. For example, TruffleHog computes the entropy of ``2b95710rD1e6287e69Z8f2E24373449d879b70c7601B3x9'' and ``ThisIsAReallyLongString'' as 4.08 and 4.11 respectively, thus having higher entropy score for the latter~\cite{tf-shannon-entropy}. As a result, the dummy string is termed as a secret. We also observed substantial instances of GitHub commit ids, such as ``0e2b3d4e3dec5f38ae95f62519eb2736f73c0b'', outputted as secrets because of ineffective entropy calculation.

\textit{\underline{3. Insufficient Filters/Prefix Regex:}} We observe that tools apply filters for HTML attributes and CSS selectors. For example, Repo-supervisor applies regex to prevent false positives such as ``input[val=`test']'' and ``button[value=`submit']''~\cite{repo-sup-css-filters}. However, the filters are insufficient as we observed strings such as ``shape=rect;rounded=1'' and ``child\{margin-bottom:10px;\}'' are still marked as secrets. In addition, tools apply prefix regex to ensure that at least one of the specified keywords related to the API key and token are within some characters (e.g., 40 characters) of the capturing group. For example, if a Strava API key is found by regex ``\texttt{[0-9a-z]\{40\}}'', then the specified prefix regex checks whether the keyword ``strava'' is present within 40 characters of the capturing group~\cite{strava-prefix-regex}. However, checking with prefix keywords does not always prevent false positives. For example, TruffleHog applies regex \texttt{((?:glpat|)[a-zA-Z0-9\-=\_]\{20,22\})} with ``gitlab'' as prefix keyword to identify GitLab tokens. However, for a string such as ``https://docs.gitlab.com/gitlab-basics/add-file.html\#add-a-file-using-the-command'', TruffleHog treats ``add-a-file-using-the-'' as a token since the string matches the regex and the prefix keyword is present within 40 characters. 

\textbf{\underline{Analysis of False Negatives:}} Since we observed a low recall score by the tools, we inspected a random sample of 50 false negatives from each tool. Below, we discuss the reasons behind the low recall score.

\textit{\underline{1. Faulty Regex:}} We observe that tools miss secrets because of employing faulty regular expressions. For example, Whispers employ regex (\texttt{.*[A-Za-z0-9\-\_]+(key|token)\$}), which expects a secret will have a ``key'' or ``token'' word at the end. However, the ``key'' or ``token'' word can be present at the start of the context of the secret (\texttt{api\_key="xxxx"}) or even not present at all, thus unable to capture secrets.

\textit{\underline{2. Skip Specific File Types:}} We observe tools skip specific file types while scanning. For example, ggshield does not scan HTML files to prevent false positives~\cite{ggshield-quick-start}. However, we observed that secrets are present in the HTML files either inside the HTML tags or in the JavaScript code embedded in the HTML files in a \texttt{<script></script>} tag. In addition, the HTML file type is in the top five file types containing secrets in the benchmark dataset (Table~\ref{secret-file-types}).

\textit{\underline{3. Insufficient Ruleset:}} We observe that tools do not have sufficient rulesets for all secret types. For example, TruffleHog does not have detectors for IGDB~\cite{igdb-api} and Mashape API~\cite{mashape-api} keys. As a result, since TruffleHog matches prefix keywords for a specific key, these API keys are not captured. We also observe that tools do not periodically add/update rules for detecting secrets. For example, the rules of the tools such as Whispers were last updated on August 25, 2021.

\textbf{\underline{False Positive Secrets Dataset:}} We created a dataset of the false positives reported by the tools to expedite the research on improving the accuracy of the tools. The dataset is stored as a relational structured data in Google
BigQuery (Dataset ID: dev-range-332204.fpsecretbench), and users can run SQL queries to access the dataset. However, the dataset may contain sensitive information, such as mislabeled true positives since the applied string-matching algorithms may mislabel the tool-reported secrets (Section~\ref{FilterCompareToolAlerts}). As a result, we will distribute only to fellow researchers and tool developers who should email the authors to access the dataset~\cite{fp-secretbench}.

\subsection{RQ2: What features are offered by the secret detection tools to aid in preventing secrets exposure?}\label{RQ2}

Tools provide features to aid developers in preventing the exposure of secrets. We categorized the features into seven categories. Table~\ref{tools-feature} presents the features offered by each tool, which we discuss as follows.

\textbf{F1: Pre-commit Hook Integration:} Pre-commit hook is a VCS mechanism that can be used for any validation before a commit is pushed. Secret detection tools can be integrated into a pre-commit hook to prevent leaking secrets. The tools will scan the source code of the current commit and reject the commit if any secret is found. Developers can employ this feature in accordance with ``shifting left'' on security~\cite{shiftleft-security}.

\textbf{F2: CI/CD Integration:} Secret detection tools offer integration with continuous integration and continuous deployment (CI/CD) pipelines such as GitHub Actions~\cite{github-actions}, Travis~\cite{travis}, and CircleCI~\cite{circleci}. As a result, if a secret is found in the deployment package, the deployment can be rejected.

\textbf{F3: Custom Rule:} Tools support adding custom rules, thus allowing developers to devise rules to detect known secrets. Tools allow adding custom detectors using regex or keywords for scanning secrets. In addition, tools support custom rules for ignoring secrets. If a dummy secret is knowingly committed in the source code, developers can devise rules to ignore that secret to reduce the false positive warnings from the tools.

\textbf{F4: Secret Verification:} If any potential secret is detected, the tool verifies the validity of the secret by calling the endpoint provided by the respective API vendor to reduce false positives. For example, TruffleHog's AWS credential detector~\cite{tf-aws-detector} performs a ``GetCallerIdentity'' API call against the AWS API to verify if the credential is active. In addition, if the secret is validated, GitHub Secret Scanner notifies the repository administrators and owners through email.

\textbf{F5: Remediation Steps:} Tools provide remediation workflows when a secret is detected to revoke and rotate the secret quickly. Tools assign the detected secret to the developer who leaked the secret. The developer can resolve the secret alert either by revoking the secret or marking it as a false positive. Tools also use developer feedback to improve their algorithm to reduce false positives. In addition, tools also provide suggestions, such as removing the secret from Git history and reviewing access logs to nullify the threat completely.

\textbf{F6: Infrastructure as Code (IaC) Script Scan:} Scanning for secrets in IaC script is essential as Rahman et al.~\cite{rahman2019seven} identified hard-coded secret is the most occurring security smell within IaC scripts. SpectralOps and ggshield provide support for scanning secrets in IaC scripts.  

\textbf{F7: Non-source Code Scan:} Developers can expose secrets in screenshots added as images in a repository and non-searchable PDFs shared for tutorials. These secrets can not be captured using regular regex matches. However, Commercial X employs Object Character Recognition (OCR) to detect secrets in images and non-searchable PDFs.

\begin{table*} [!htb]
\footnotesize
\centering
\caption{Seven categories of features and additional secrets metadata provided by each tool.}
\label{tools-feature}
    \includegraphics[width=\textwidth]{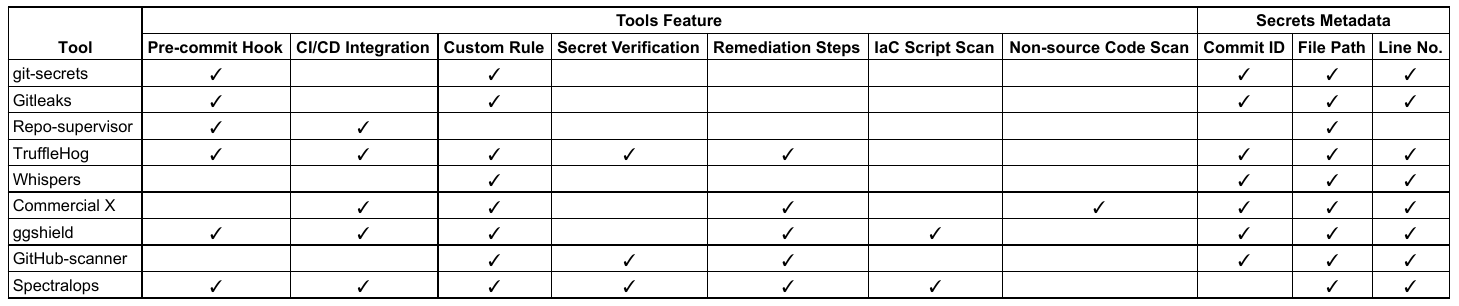}
\end{table*}

\section{Discussion and Recommendations} \label{Discussion}
Below we discuss our findings and make recommendations.

\textbf{Developers should employ tools based on the type of secrets present in the project.} Table~\ref{tool-accuracy} shows that tools miss secrets as the recall (Case 2) varies between 17\% and 88\%. However, if developers know the secret types present in the project, selecting tools based on secret types can yield higher recall. For example, for ``Database Server and URLs'' category, the recall (Case 2) score of TruffleHog is 98\% (Table~\ref{recall-secret-categories}), whereas the overall recall (Case 2) score is 52\% (Table~\ref{tool-accuracy}).

\textbf{Tool vendors should update detection rules periodically.} According to the State of APIs Report from Rapid~\cite{rapid-state-of-api}, API types are expanding, and API adoption is on the surge, with 63\% of developers relying more on APIs in 2022. However, we observe that tools do not update the detection rules for API keys and tokens. For example, the rules of Whispers were last updated on August 25, 2021. We recommend tool vendors to update detection rules periodically to prevent missing secrets.

\textbf{Tool vendors should correctly employ secret verification by collaborating with API vendors.} We find that tools verify the found secrets with the API endpoints (F4). As a result, tools show relatively higher precision by reducing false positives. For example, before the verification option was enabled (\texttt{--only-verified}), TruffleHog's precision was 6\%, outputting almost 100K alerts for our benchmark. In contrast, the precision changed to 90\% when the verification was enabled and outputted only 611 secrets. However, verification methods are not 100\% correct as we observe 10\% false positives. For example, the tool tagged dummy server URLs such as ``http://dyn.example.com:password@dyn.dns.he.net" as secrets. In addition, TruffleHog does not report an active secret if the API endpoint is unreachable~\cite{tf-endpoint-unreach}. We also find that GitHub has a secret scanning partner program~\cite{github-partner-program} where API vendors can join in scanning their API keys and tokens in GitHub repositories and receiving notifications for quick remediation. However, only 66 API vendors have joined the program~\cite{github-partner-patterns}. Therefore, we recommend that API vendors collaborate with tool vendors in correctly employing secret verification to prevent the exposure of secrets.

\textbf{Tool vendors should develop automated technology to revoke and rotate secrets as remediation steps quickly.} We find that tools provide remediation workflows when a secret is detected (F5). However, currently, the workflow is a manual process where the leaked secret is assigned to the developer to revoke and rotate the secret. In addition, developers have to sanitize the Git history by themselves using history sanitizing tools such as BFG repo-cleaner~\cite{bfg-repo-cleaner}. %As a result, the process takes a longer time to complete. 
However, recent research~\cite{secret-leak-one-minute} shows that malicious actors take only one minute to start making calls with the leaked API keys. Therefore, we suggest that tool vendors develop an automated workflow that the organization can employ in their system. The organization can mark the used secrets, and if a secret is reported that are among the used secrets, the workflow will automatically revoke and rotate the secrets. In addition, the workflow will sanitize the Git history without developers' manual effort, deploy new artifacts if needed, and review access logs to find any breaches.

\section{Ethics and Data Protection} \label{Ethics}
Since the dataset of false positive secrets may contain mislabeled true positives, we will distribute the dataset selectively. To prevent unethical use, researchers and tool developers will sign a data protection agreement with us. Following that, we will use their email addresses to grant them access to our dataset from Google BigQuery.  In addition, we have redacted/obfuscated example secrets presented in our paper.

\section{Threats to Validity} \label{ThreatToValidity}
In this section, we discuss the limitations of our paper. 

\textit{\uline{Tool Selection}}: Our study's list of tools is not exhaustive. Though we have chosen the tools based on the selection criteria mentioned in Section~\ref{selectionoftools}, we could not access proprietary tools such as Cycode~\cite{cycode} and CredScan~\cite{credscan}. As a result, we do not claim the findings we have in Section~\ref{Results} to be generalizable for all tools.

\textit{\uline{Benchmark Dataset}}: Our selection of benchmark dataset is susceptible to bias. Basak et al.~\cite{secretbench} curated SecretBench using open-source tools Gitleaks and TruffleHog, which also poses bias to the result of these two tools. However, they manually inspected and labeled each extracted secret using the tools. Out of 97,479 reported secrets of these two tools, 15,084 are true secrets. We used the true secrets to compare the tools of our study. SecretBench also has the drawback of only extracting secrets from GitHub repositories rather than from other VCSs, such as GitLab and BitBucket. Since SecretBench is the only publicly-available dataset, we could not compare the tools with another benchmark dataset to mitigate the potential bias.

\textit{\uline{Secrets Matching}}: We employed two string matching algorithms, Jaro-Winkler Similarity, and Gestalt Pattern Match, to match a secret with the benchmark for some tools. The similarity cut-off scores for both the algorithm we chose poses a threat to internal validity. However, we randomly selected 100 unique reported secrets from each tool and found that the combination of both algorithms' cut-off scores correctly labeled 97\% of the secrets.

\textit{\uline{Precision for Each Secret Category:}} We have the category of a secret and the number of secrets in a category of benchmark dataset. As a result, we could calculate the recall of each category by checking if the secrets of the specific category of the benchmark are present in the tool-reported secrets. However, we could not calculate the precision for a category since the tool can output false positives, which requires manual inspection for categorization.

\section{Related Work} \label{RelatedWork}
The root causes of the widespread leakage of secrets in software artifacts have been studied in prior work~\cite{meli2019bad,rahman2019share,rahman2019seven, igibeksecret, rahman2021different}. Researchers have found that the most prevalent insecure practice adopted by developers causing secret leakage is hard-coded secrets in software artifacts. In 2019, Meli et al.~\cite{meli2019bad} studied a 13\% snapshot of public GitHub repositories and found over 100K hard-coded secrets in the source code. Within Infrastructure as Code (IaC) scripts, Rahman et al.~\cite{rahman2019seven} studied a recurring coding pattern known as ``security smells" which are indicators of security flaws that can result in potential security breaches. They investigated 5,232 IaC scripts extracted from 293 open-source repositories. They found seven security smells and the hard-coded secret is the most occurring security smell with 1,326 occurrences. In addition, hard-coded secrets have also been found in GitHub Gists that are used to share code snippets among developers. Rayhanur et al.~\cite{rahman2019share} investigated 5,822 publicly available Python Gists and found 689 instances of hard-coded secrets in the code snippets. All of these prior works suggest that hard-coded secrets have been leaking in different forms in software artifacts.

To prevent secret leakage in software artifacts, researchers have suggested developers follow secure practices for secret management~\cite{basaksecretpractice, basak2023challenges}. Basak et al.~\cite{basaksecretpractice} conducted a grey literature review of Internet artifacts, such as blog articles, and identified 24 practices comprised of both developer and organization practices. They suggested using VCS scan tools to prevent accidental commit of secrets. In another work, Basak et al.~\cite{basak2023challenges} investigated the questions related to checked-in secrets in Stack Exchange (SE) and the solutions posted by the SE users to mitigate the challenge. They identified that the SE users have also suggested using VCS scan tools to prevent accidental secrets leakage. However, in 2021, Rahman et al.~\cite{rahman2022secret} conducted a developer survey in XTech company (Anonymized) and found that developers bypass the alerts of scan tools as the tools generate a lot of false positives. Recent research~\cite{9794113,saha2020secrets,10063545} utilizes ML algorithms to reduce false positives in secret detection. Saha et al.~\cite{saha2020secrets} employed a Voting Classifier (a combination of Logistic Regression, Naive Bayes, and SVM) to distinguish real secrets from false positives. Feng et al.~\cite{9794113} applied deep neural networks to uncover the intrinsic characteristics of textual passwords and detect real passwords by reducing false positives.

At present, many open-source and proprietary secret detection tools are available. However, developers face difficulty choosing one tool out of many because of a high number of false positives. As far as we know, no research has been conducted yet evaluating and comparing the existing secret detection tools. In this work, we concentrated our research efforts on evaluating and comparing 9 secret detection tools.

\section{Conclusion} \label{Conclusion}
We investigated five open-source and four proprietary secret detection tools against a benchmark dataset containing 818 GitHub repositories. We found that the top three tools based on precision are: GitHub Secret Scanner (75\%), Gitleaks (46\%), and Commercial X (25\%), and based on recall are: Gitleaks (88\%), SpectralOps (67\%) and TruffleHog (52\%). We also provided tools performance based on secret type to aid developers select the best tools for their use cases. Our manual analysis of the reported false positives indicates that generic regex and ineffective entropy calculation are the reasons for high false positives. We also analyzed the false negatives and found that faulty regex, skipping file types, and insufficient rulesets for secret detection are the reasons for low recall. In addition, we provided a dataset of false positives to expedite the research in secret detection. We also categorized the features offered by the secret detection tools to aid in preventing the exposure of secrets. We recommend developers choose tools based on secret types present in the project to prevent missing secrets. In addition, we recommend future research on developing an automated technology for quick remediation of the exposed secret.


\section*{Acknowledgment}

This work was supported by the National Science Foundation (NSF) 2055554 grant. The authors would also like to thank the Realsearch research group for their valuable input on this paper.
%for later:  grant number 2055554.



\bibliographystyle{IEEEtran}
%\bibliographystyle{plain}
\bibliography{bibliography}

%\bibliographystyle{plain}
%\bibliography{bibliography}

% \section*{Appendix}
% \input{Tables/QuestionCountPerYear}
% \input{Tables/tagskeywords}

\end{document}